\documentclass[sigplan,screen]{acmart}

\AtBeginDocument{%
  \providecommand\BibTeX{{%
    \normalfont B\kern-0.5em{\scshape i\kern-0.25em b}\kern-0.8em\TeX}}}

\setcopyright{none}

\acmConference[]{2021}
\acmBooktitle{under review}
\acmPrice{}
\acmISBN{}

\usepackage{booktabs}
\usepackage{makecell}
\usepackage{multirow}

\newcommand{\SYS}{DISC}
\newcommand{\NAME}{\textit{\SYS} }

\newcommand{\parahead}[1]{\textbf{\textit{#1}}}

\begin{document}

\title{\NAME: A Dynamic Shape Compiler for Machine Learning Workloads}

\author{Kai Zhu, Wenyi Zhao, Zhen Zheng, Tianyou Guo, Pengzhan Zhao, Feiwen Zhu} \author{Junjie Bai, Jun Yang, Xiaoyong Liu, Lansong Diao, Wei Lin}
\affiliation{\country{Alibaba Group}}
\affiliation{\country{\{tashuang.zk, kevin.zwy, james.zz, tianyou.gty, pengzhan.zpz, feiwen.zfw\\j.bai, muzhuo.yj, xiaoyong.liu, lansong.dls, weilin.lw\}@alibaba-inc.com}}


\begin{abstract}

Many recent machine learning models show dynamic shape characteristics. 
However, existing AI compiler optimization systems suffer a lot 
from problems brought by dynamic shape models, 
including compilation overhead, memory usage, optimization pipeline and deployment complexity.
This paper provides a compiler system to natively support optimization for dynamic shape workloads, named \NAME.
\NAME enriches a set of IR to form a fully dynamic shape representation. 
It generates the runtime flow at compile time to support 
processing dynamic shape based logic, 
which avoids the interpretation overhead at runtime 
and enlarges the opportunity of host-device co-optimization.
It addresses the kernel fusion problem of dynamic shapes 
with shape propagation and constraints collecting methods.
This is the first work to demonstrate how to build an end-to-end 
dynamic shape compiler based on MLIR infrastructure.
Experiments show that \NAME achieves up to 3.3$\times$ speedup than TensorFlow/PyTorch, 
and 1.8$\times$ than Nimble.

\end{abstract}


\acmYear{2021}\copyrightyear{2021}
\setcopyright{acmcopyright}
\acmConference[EuroMLSys '21]{The 1st Workshop on Machine Learning and Systems}{April 26, 2021}{Online, United Kingdom}
\acmBooktitle{The 1st Workshop on Machine Learning and Systems (EuroMLSys '21), April 26, 2021, Online, United Kingdom}
\acmPrice{15.00}
\acmDOI{10.1145/3437984.3458838}
\acmISBN{978-1-4503-8298-4/21/04}
\maketitle

\section{Introduction}
\label{sec:introduction}

Machine learning technique evolves fast in recent years. 
It is applied in a wide range of areas, such as image/speech recognition, 
translation, recommendation to serve people's life.
One reason of the boost of machine learning is the growth of computing power.
With the help of machine learning frameworks (TensorFlow\cite{abadi2016tensorflow}, PyTorch\cite{paszke2019pytorch}, MXNet\cite{chen2015mxnet}), 
machine learning algorithm representation can be mapped to powerful devices for convenient execution. 
However, this mapping is non-trivial and there is still a performance gap, 
especially for new models and scenarios.

The recent developed models expose dynamic shape problem, which is less optimized with current techniques.
The operations suffering most from dynamic shape problems are those with small sized computations, 
like element-wise and reduction operations. 
Traditional techniques, like XLA\cite{xla}, usually apply fusion approach to 
reduce the off-chip memory access and frequent kernel launch overhead for such operations. 
However, the existing kernel fusion engines could only generate kernels 
with static shape information inferred during compilation time.
This results in a problem that, these fusion engines will compile and generate kernel for every emerging shape,
even though some of them share the same computation pattern.
It leads to severe compilation overhead when the number of shapes is large.
Due to this reason, XLA is usually closed for dynamic shape workloads to prevent negative optimization.

Note that large ops, like GEMM/Conv, do not suffer much from dynamic shapes 
as they usually go through library calls (cuDNN, cuBLAS, oneDNN) rather than compilation optimizations.
We focus on small sized ops optimization targets in this paper.

There are some workaround solutions for dynamic shape problem based on XLA. 
Developers can only cluster ops that have static shape for XLA to optimize, 
and leave ops with dynamic shape features run without fusion. 
This loses optimization opportunities to a extent. 
Furthermore, some workloads only have dynamic shaped ops in practice.
Another workaround is to form tensors into a specific shape with padding and slicing, 
which introduces redundant computations and may lead to negative optimizations.
None of the workarounds solves this problem fundamentally.

MLIR\cite{lattner2020mlir} provides the infrastructure towards a new machine learning compiler.
It brings high extensibility to new functions and compatibility to existing optimization buildings.
Meanwhile, it naturally supports dynamic shape optimization with its design philosophy.
However, what it brings is the infrastructure, but not the solution to dynamic shape problem itself.
Nimble\cite{shen2020nimble} is a compiling framework based on TVM to address dynamic shape problem, 
which is a concurrent work with \NAME and \NAME has an earlier RFC release\cite{discrfc}. 
It provides a compiler framework capable of adapting to dynamic shaped ops at runtime. 
The runtime control logic is pre-built as a VM component. 
A problem of Nimble is that, it pre-builds runtime control as VM, 
which loses the opportunity to explore host-device co-optimization.
Meanwhile, the VM approach brings interpretation overhead.

We propose \NAME, a \underline{d}ynam\underline{i}c \underline{s}hape 
\underline{c}ompiler for machine learning workloads.
We build \NAME based on MLIR infrastructure to leverage 
its native support of dynamic shape from high level design perspective. 
\NAME tackles several main problems of dynamic shape optimization. 

The first is the lack of a complete representation of dynamic shape computations with existing IR. 
Note that MLIR does not provide dynamic shape IR expression directly. 
We do not build a set of new IR from scratch, but introduce \textit{DHLO} based on HLO dialect, the IR already used in XLA. 
This approach enables us to reuse some existing building blocks of XLA and \textit{MLIR-HLO dialect}.

The second is to build efficient runtime flow to support dynamic shape logic.
Instead of building a VM to interpret dynamic shaped flow at runtime, 
we generate the code of runtime flow just-in-time at compile time. 
This avoids the interpretation overhead of a VM.
Meanwhile, this approach enlarges the opportunities of host-device co-optimization 
as \NAME compiles the device computation and host-side logic all together.

The third is to generate efficient fusion kernels without knowing full shape information.
We check the shape compatibility of two ops with two collected characteristics.
We first make use of shape propagation property between producers and consumers to fuse adjacent ops. 
Furthermore, we collect shape constraints when lowering computation graph to \textit{DHLO}. 
The extra shape constraints information allows us to form larger scope of fusion 
to further reduce off-chip memory access and kernel launch overhead.

Finally, \NAME supports multiple machine learning frameworks (TensorFlow and PyTorch) 
with the hub of \textit{DHLO}. 
Meanwhile, \NAME supports the mix of static and dynamic optimization.
When \NAME finds a sub-graph with static shape, it will fallback to static optimization for better performance.

Experiment results show that \NAME outperforms TensorFlow/PyTorch 
with 2.27$\times$ speedup in average for 6 popular models, 
and Nimble with 1.8$\times$ speedup for \textit{transformer}.

This paper has the following main contributions.

\begin{itemize}

\item It is the first work to demonstrate how to build a compiler supporting dynamic shape efficiently with MLIR infrastructure.

\item It proposes an approach to support dynamic shape processing with  
the design of fully dynamic IR and compile-time generated runtime flow.

\item It addresses the fusion problem without full shape information, 
specifically with the additional shape constraints collecting method.

\item It supports multiple machine learning frameworks and the mix of static/dynamic optimization. 

\end{itemize}

\section{Background}
\label{sec:background}

The computation graphs of modern machine learning models consist of both compute and memory intensive ops.
In this paper, we refer to GEMM and Conv as compute intensive op and other ones as memory intensive op.
Compute intensive ops are usually called with pre-built libraries, like cuDNN and cuBLAS, 
in popular machine learning frameworks.
While memory intensive ops are optimized with AI compilers with kernel fusion and code generation techniques.
Note that a single memory intensive op, like an \textit{Add} op, is too light weighted to build a library for it. 
Meanwhile, the combination of memory intensive ops varies in different workloads 
and it is infeasible to pre-build fused kernels for such ops.

\parahead{Static Shape Oriented Compiler.}
We take XLA\cite{xla}, state-of-the-art compiler optimization engine for memory intensive ops, 
to explain how a static shape compiler works. 
Given a computation graph, XLA firstly translates it into HLO IR. 
It then finds ops that can be fused together and generates fusion kernels, 
which will be cached according to fusion pattern.
The fusion pattern contains op sequence with full shape information.
When XLA meets a fusion pattern, it will first check whether this pattern is already cached.
It will use the binary directly if hit, otherwise it will compile for the new pattern and cache the compiled result.

\parahead{Lack of Attention for Dynamic Shape.}
The process of XLA works well for static shape scenario, but is less efficient for dynamic shape workload.
A typical case is Seq2seq models with varying input/output sequence length.
Although the computation graph does not change, 
XLA needs to re-compile for the fused kernels for samples with different length.
When the number of shapes is large, the overhead of compilation time 
and host/device memory usage to cache makes static shape oriented compilation not usable.
Other typical workloads suffering from dynamic shape issues includes 
CV workloads processing different image sizes (like object detection), 
and sparse workloads with \textit{Unique}\cite{tfunique} ops generating output tensors with varying shapes.

\NAME generates fused kernels adaptive to any coming shapes and avoids re-compilation.
The basic insight is that we do not need to consider shape information 
to check whether two fusion patterns are the same for code generation.
Note that \NAME only targets dynamic shapes with static rank, 
as we do not find dynamic rank a popular behavior.

\parahead{MLIR Infrastructure} We build \NAME based on MLIR infrastructure\cite{lattner2020mlir}, 
which aims to help building reusable and extensible compiler infrastructure.
We choose MLIR as it is open for extension and could accommodate existing optimizations 
based on other IRs with dialect conversion.
Specifically, it allows to reuse the existing optimization components of XLA 
by lowering \textit{MLIR-HLO dialect} to \textit{HLO}.

However, what MLIR itself offers is a flexible infrastructure, 
rather than a solution to problems such as the optimization of dynamic shape ops. 
\NAME shows a way to build a complete optimization system that targets dynamic shape workloads with MLIR.

\section{Overview of \NAME}
\label{sec:ovewview}

\begin{figure}
    \includegraphics[scale=0.4]{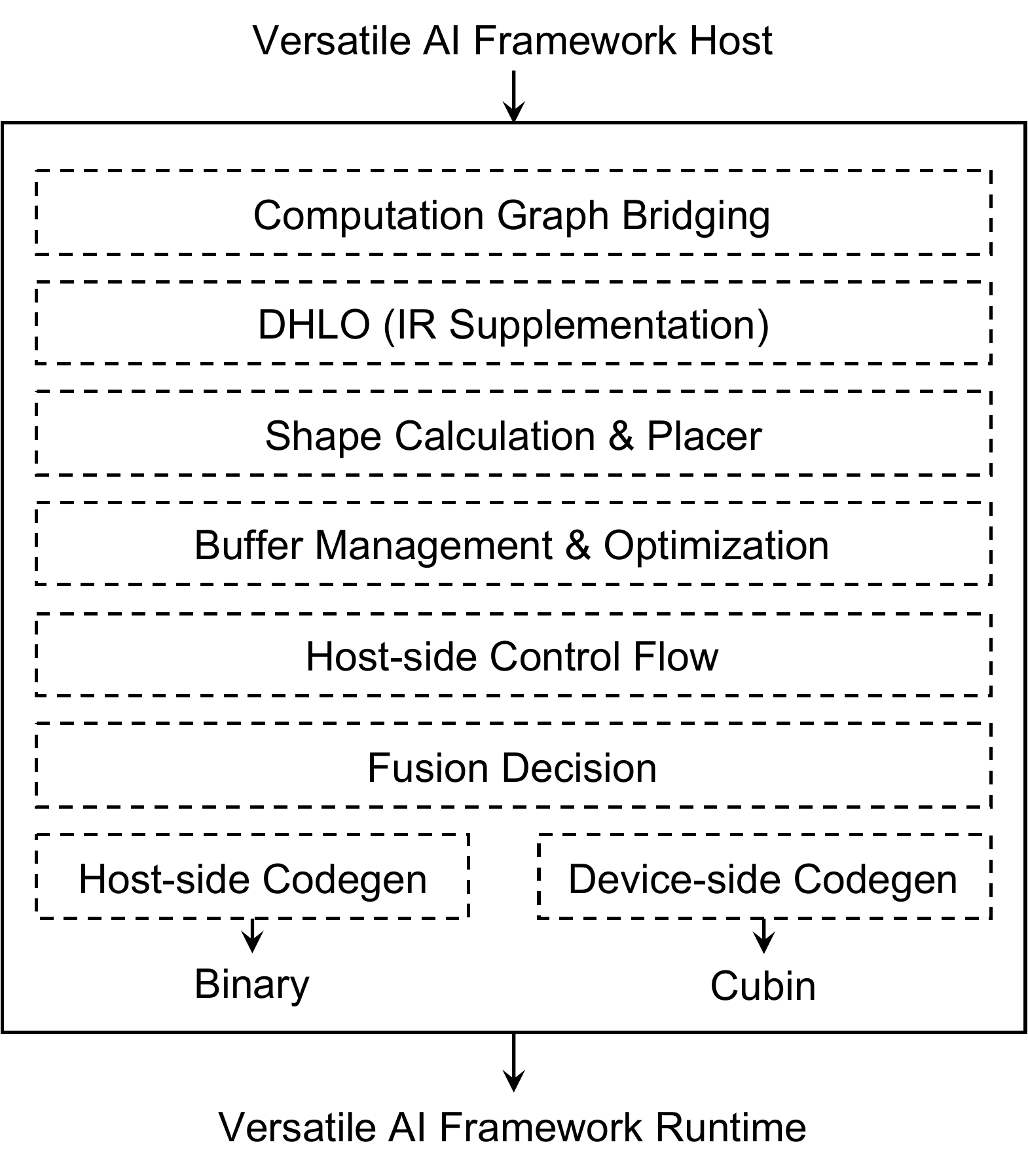}
    \caption{\NAME framework overview.}
    \label{fig:disc}
\end{figure}

Figure \ref{fig:disc} describes the overview of \NAME framework.

The first component of \NAME is \textit{computation graph bridging}, 
which lowers the computation graph described with versatile AI frameworks 
into the hub IR used in \NAME (section \ref{subsec:dhlo}). 
\NAME also collects shape constraint information in this layer 
to help with fusion optimization (section \ref{subsec:shape-infer}).

The hub IR, \textit{DHLO}, is extended from \textit{HLO} dialect to support fully dynamic shape.

The basic execution flow for dynamic shape workload is to 
compile without full shape information and get shape information at runtime.
\NAME separates shape computation and data processing
during compilation.
It complies and code-gen the \textit{shape calculation} logic.
The \textit{placer} component places \textit{shape calculation} logic on host side 
and tensor computation kernels on device side.
The generated shape inference function will be processed on host side when executing models.

The \textit{buffer management} component manages buffer lifetime of tensors in computation graph.
\NAME generates the code about buffer allocation, reuse and free logic at compile time, 
and executes the compiled flow at runtime.
The basic optimization rule is to free buffer as soon as it has no users, 
and reuse buffers as much as possible according to "shape compatibility".

The \textit{host-side control} is responsible for external library lowering, 
kernel launch management, device management, 
and the interaction between compiler engine and AI framework hosts.
Similar with other runtime logic, \textit{host-side control} is also generated at compile time.
This design is to prevent the interpretation overhead of previous works (section \ref{subsec:runtime-gen}) 
and enrich the opportunity of host-device co-optimization.

\textit{Fusion decision} relies on op schedule compatibility analyzing.
\NAME decides to fuse ops according to shape hints, 
which comes from shape propagation and shape constraints analysis.

The \textit{host-side codegen} and \textit{device-side codegen} component 
generates binary for host side and CUDA cubin for GPU side.

Finally, \NAME plugins the compiled and optimized computation graph into AI frameworks.
It adapts to the runtime implementations of versatile AI frameworks,
with a small set of interface APIs to handle compiler-framework interactions.

\section{System Design}
\label{sec:design}

The first problem of a dynamic shape compiler is that, it lacks an IR expression.
We extend \textit{HLO} and introduce \textit{DHLO} as the IR to support full dynamic shape features (\ref{subsec:dhlo}).
With \textit{DHLO} as the hub IR, \NAME could support multiple front ends and backends (\ref{subsec:multi-framework}).
To meet the runtime requirement of dynamic shape supporting, 
\NAME generates the runtime flow (\ref{subsec:runtime-gen}) to avoid the interpretation overhead.
Finally, \NAME analyzes the shape hints to support fusion optimization (\ref{subsec:fusion}).

\subsection{DHLO: IR Supplementation}
\label{subsec:dhlo}

MLIR infrastructure is flexible and easy to extend to support versatile features.
However, it only provides the infrastructure but not specific IR design to support dynamic shape directly.
We choose \textit{HLO} IR to build MLIR dialet for dynamic shape problem, 
because \textit{HLO} already supports many op descriptions and different frameworks.
However, \textit{HLO} is designed for static shape compiler optimization and 
lacks expressing ability for dynamic shape in some cases.
As a solution, we extend \textit{HLO} with a set of IR supplementation and introduce \textit{DHLO}.

\begin{figure}
    \includegraphics[width=\columnwidth]{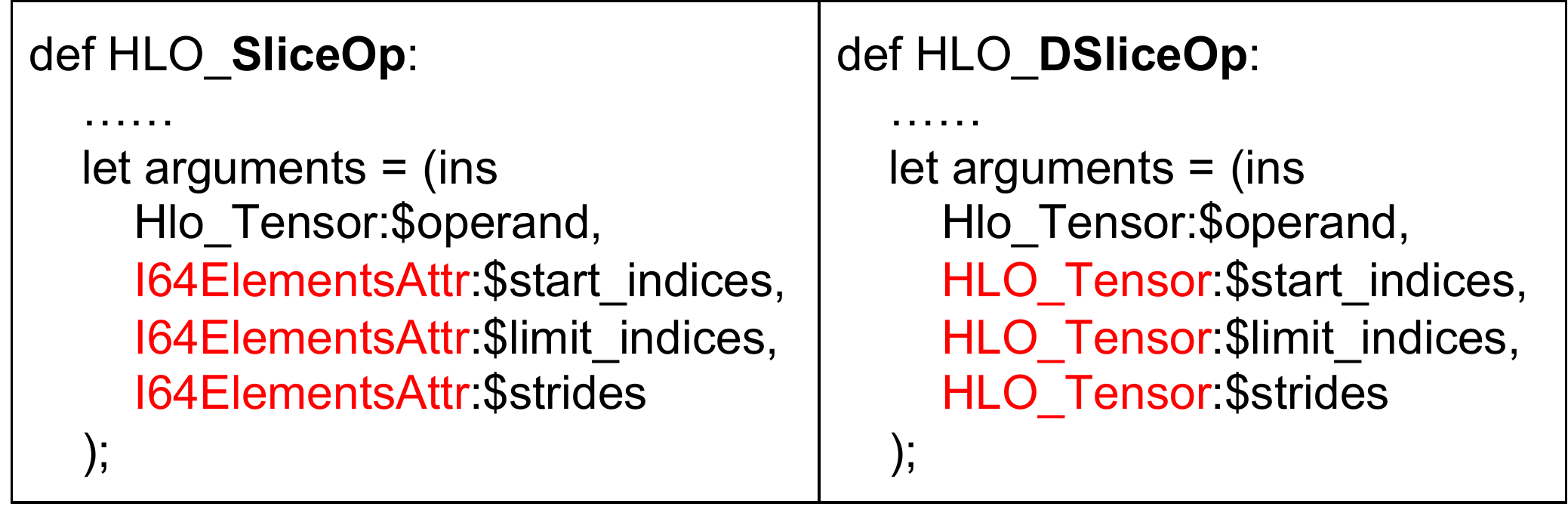}
    \caption{Static \textit{slice} op (left) and dynamic \textit{slice} op (right).}
    \label{fig:dslice}
\end{figure}

The insight of IR supplementation is to replace compile time constant folding to runtime tensor dataflow.
Specifically, the target ops for which to extend IR representation is 
those with attributes being constant folded in \textit{HLO}, like \textit{slice}, \textit{pad}, \textit{broadcast}, et al.
In \textit{DHLO}, we replace the constant attributes with tensor arguments to support dynamic shape behavior.
Take \textit{slice}\cite{slice} as an example, as is shown in figure \ref{fig:dslice}.
A \textit{slice} op extracts a sub-tensor from the input tensor given indices of the bounding box.
The indices of the bounding box  are constants at compile time in \textit{HLO} 
(\textit{start\_indices}, \textit{limit\_indices}, and \textit{strides} shown in figure \ref{fig:dslice}).
However, these indices vary in shapes for dynamic shape workloads 
and constant folded expression is infeasible.
Instead, we define the indices of the bounding box as tensor arguments of \textit{slice}.
Such extension works well for dynamic shape scenarios as tensor value are generated at runtime.

Note that \textit{DHLO} is only an extension but not to replace the whole \textit{HLO}. 
Since for many ops, like elementwise \textit{Add/Mul} etc, their definition in \textit{HLO} has enough expressiveness  
to support dynamic shape and we leave them as they are.

\subsection{Generated Runtime Flow}
\label{subsec:runtime-gen}

One challenge of dynamic shape compiler is that, 
compiling is a static action, while we aim to handle dynamic behavior.
Current state-of-the-art compiler optimization engines, like XLA\cite{xla}, 
require to know all shape information before compiling.
XLA generates code and builds buffer management and kernel launch sequence statically at compile time.
This compiling process does not work without shape information known ahead.

A dynamic shape compiler needs to generate code that is adaptive to handle any shapes encountered at runtime.
Nimble\cite{shen2020nimble} designs a VM to interpret runtime flow of graph computation with dynamic shape.
It interprets tensor shapes and organizes runtime logic adaptively.
Rather than using a interpreter, \NAME compiles and generates the code of computations on both host and device side, and also runtime flows (buffer management, kernel launch, et.al.).
The compiler approach of runtime flow reduces the overhead of VM interpretation in Nimble.

\subsubsection{Adaptive Shape Inference}
\label{subsec:shape-infer}

The shape inference component in \NAME has two stages.
The first is to identify the shape constraints at compile time 
without knowing specific shape value. 
The second is to emit runtime codes to calculate specific shape values given input tensor. 
The former serves for code generation optimization and the latter serves for correct execution.

\parahead{Shape constraints.}
Without concrete shape value when compiling, we lose some optimization opportunities. 
This is a common problem of current dynamic shape related compiler techniques.
\NAME reveals that, we can still obtain some additional shape constraint information 
to help generating efficient kernels.

\NAME explores two kinds of shape constraints. 
The first one is called dimension size equality constraint.
This kind of constraint reveals whether one dimension size of a tensor is equal to another
dimension of the same tensor or any dimension of another tensor.
The second one is called tensor size equality constraint, which reveals whether two tensors have the same number of elements.
Such shape constraints can be useful in both IR optimization and code generation stage.  An optimization case in IR optimization stage is that when we know that two ops manipulate tensors with the same or compatible shape,  we can make decision to fuse them together. In code generation stage, these constraints enables more aggressive index calculation simplification.

\NAME collects shape constraints from two sources.
In the first case, we infer shape constraints captured by the \textit{DHLO} op semantic.
For example,  the input tensor and output tensor of a \textit{TransposeOp} should  have the same tensor size. Similarly, the input tensor and output tensor of a \textit{AddOp} should have the same shape according to op definition.
In the second case, we collect shape constraints captured by the high level ops from frameworks and inject such information into \textit{DHLO} in \textit{computation graph bridging}.
Take \textit{SplitOp} in \textit{Tensorflow} as an example.  It divides a tensor along a dimension evenly, which implies that all outputs of this op have the same shape. 
 A \textit{TF.SplitOp} will be lowered to multiple independent \textit{DHLO.SliceOp}, which actually have the same shapes. However such kind of information is lost after being lowered to \textit{DHLO} without explicit shape constraint.


\parahead{Shape calculation.}
Different from static shape compilers that only needs to generate code for 
computations themselves with constant folded shape information, 
\NAME generates the code of shape inference and kernel computation sub-graphs separately.
Shape calculation computation is light weight and \NAME place it on host side (CPU),
while the sub-graph manipulating tensors are placed on device side (GPU).
The placement logic is similar with Nimble\cite{shen2020nimble}. 
The difference is that, \NAME applies compiling approach to generate the code of 
computation, shape inference and placement logic all together,
rather than using a pre-built VM interpreter for runtime control like in Nimble.
This avoids extra interpretation overhead.
Meanwhile, this approach brings opportunities of joint optimizations 
between host and device sides.


\subsubsection{Dynamic Buffer Management}

With emitted codes calculating shapes of each buffer at runtime, 
\NAME is able to manage the buffer dynamically by emitting \textit{alloc} and \textit{dealloc} instructions. 
For the considerations of buffer optimization which aims at reducing the buffer allocation overhead, 
we apply two approaches: 
1) Based on shape constraint in the IR, performing buffer liveness analysis and optimization;
2) Lowering the \textit{alloc} and \textit{dealloc} with a cached allocator, which is the allocator provided by \textit{TensorFlow/PyTorch} in our case. 


\subsubsection{Host-side Control}
Host-side codes are emitted in a unified compiler flow, making it possible for a joint optimization in the consequent passes. Besides shape calculation, it also includes launch dimension calculation, kernel launch, vendor library calls and device management instructions such as initialization, synchronization, cubin loading etc.

\subsection{Fusion and Code Generation}
\label{subsec:fusion}

Kernel fusion of memory-bound ops is one of the main optimizations of current AI compilers. 
A common fusion strategy is to allow memory bound ops with the same number of elements to be fused together.
However, the tensor shapes to process are not known at compile time for dynamic shape scenarios.
It is non-trivial to determine which ops could be fused together to get performance benefit.

\parahead{Shape hints collection.}
We determine whether two ops have the same shape with two hints.
The first is \textbf{shape propagation}. 
Such as the operands of an \textit{Add} op must have the same tensor shape with \textit{Add}'s consumer.
\NAME maintains a table to indicate the propagation property of each op.
Specifically, some ops may have the same shape propagation property, like \textit{Add} and \textit{Sub}.
We classify ops according to their shape propagation properties in the table to avoid repeated enumeration.
The second is \textbf{shape constraints} \NAME collects, as described in Sec. \ref{subsec:shape-infer}.

\parahead{Shape-adaptive fusion configuration.}
For fusion code generation of memory bound patterns, 
we tend to choose templates that are friendly to a wide range of shapes,
like the classical loop fusion and input fusion with reduce operation as the root.
However, there are still aspects to react differently 
with variant runtime shapes for better performance, 
like the selection of launch dimensions, the decision of whether to do loop vectorized load/store, 
and whether an implicit broadcast is necessary etc. 
For these aspects, we generate different versions of kernels,
and generate selection logic from host-side to launch a proper kernel 
at runtime for each incoming shape.

\subsection{Multiple Framework Support}
\label{subsec:multi-framework}

\NAME is able to serve multiple AI frameworks, 
like \textit{TensorFlow}\cite{abadi2016tensorflow}, \textit{PyTorch}\cite{paszke2019pytorch} et.al.
Meanwhile, it could be lowered with both static  and dynamic shape compiler.
We use \textit{DHLO} as the hub IR to connect different parts together.
This intermediate layer simplifies the adaptation.

Specifically, \NAME does not lower all computation graphs to dynamic shape compiler.
Instead, it will lower computation graphs to static shape compiler 
when shapes are known at compile time or the number of shapes is acceptable.
This is because static shape compiler engine could usually achieve better performance 
with the enriched information than dynamic shape compiler.

\subsection{Static Shape Library Support}

For compute intensive ops, different shapes may require different optimization 
to achieve the best performance.
Nimble\cite{shen2020nimble} choose to tune the kernel under a set of fixed shapes. 
The kernel is guaranteed to work on other shapes but the performance may not be the best. 
In order to balance the dynamism and performance, we implement an interface to choose the best kernel from a library according to different runtime shapes. 
The library contains both vendor libraries such as cuBLAS/cuDNN, 
and pre-generated kernels that has been hand-tuned for each shape.

\section{Evaluation}
\label{sec:evaluation}

In this section, we evaluate \NAME using a variety of machine learning applications
with different characteristics.
These workloads are realized with different frameworks on GPU, 
as is shown in table \ref{tbl:workloads-for-evaluation}. While \NAME is also applicable to
devices other than GPU.

\begin{table}
\begingroup
\caption{Workloads for evaluation.}
\label{tbl:workloads-for-evaluation}
\begin{center}
\begin{small}
\renewcommand{\arraystretch}{0.85}
\begin{tabular}{clccc}
\hline
Model                 & Framework  & Batch Size \\ \hline
\multirow{2}{*}{ASR}  & TensorFlow & 1          \\
                      & PyTorch    & 1          \\ 
Seq2seq               & PyTorch    & 64         \\ 
TTS                   & TensorFlow & 1          \\ 
BERT                  & PyTorch    & 1          \\ 
Ad Ranking            & TensorFlow & 512        \\ 
Transformer           & TensorFlow & 1          \\ \hline
\end{tabular}
\end{small}
\end{center}
\endgroup
\end{table}

To demonstrate the benefits of \NAME, we do comparison with TensorFlow/PyTorch and Nimble.
Note we do not compare with XLA, as which brings severe compilation overhead 
for these dynamic shape workloads and shows performance degradation.
We collect data on NVIDIA T4 GPU, with CUDA toolkit 10.0.

\subsection{Comparing with TensorFlow/PyTorch}

\begin{figure}
    \includegraphics[width=\columnwidth]{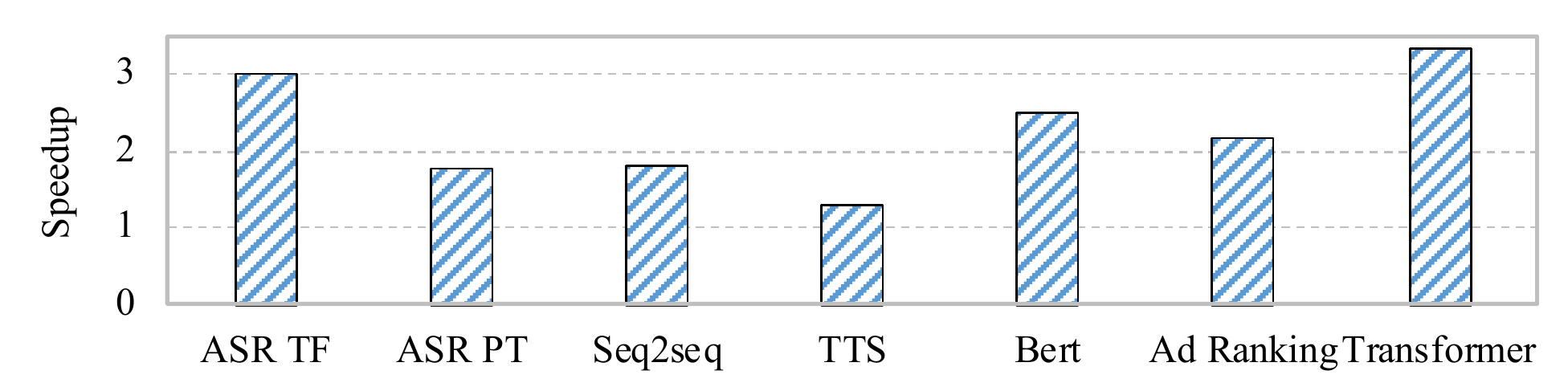}
    \caption{Speedup comparing with TensorFlow/PyTorch.}
    \label{fig:tf-pt-speedup}
\end{figure}

As is shown in figure \ref{fig:tf-pt-speedup}, \NAME achieves up to 
3.35$\times$ speedup comparing with TensorFlow/PyTorch, 2.27$\times$ in average. 
The benefit mainly comes from kernel fusion of memory intensive ops, 
which reduces off chip memory access and kernel launch overhead.
We analyze the benefit with several case studies.

\parahead{Transformer} 
We collect the breakdown information of \textit{transformer}.
We find the compute intensive ops show similar execution time for TensorFlow and \NAME version, 
while the memory intensive ops show much better performance with \NAME.
It spends 21.52ms with \NAME for memory intensive ops, 
while spending 66.06 with original TensorFlow.

One benefit comes from the reduced off-chip memory access by kernel fusion. 
The other benefit is the reduced kernel calls. 
TensorFlow results in 42884 kernel calls for memory intensive ops, 
while \NAME only has 6186 kernel calls.


\parahead{BERT} The performance speedup of \textit{BERT} also mainly comes from 
the optimization of memory intensive ops with fusion. 
The execution time of memory intensive ops is reduced from 5.96ms in PyTorch to 3.33ms in \NAME, 
and the kernel calls reduce from 198 to 97 times.

We compare \NAME with TensorRT\cite{vanholder2016efficient} realization for \textit{BERT} based on onnx-tensorrt\cite{onnx2trt} workflow, and find \NAME achieves 1.3$\times$ end-to-end speedup.
The time memory intensive ops spent drops from 4.99ms with TensorRT to 3.33ms with \NAME.


We collect breakdown for other workloads and observe similar behavior with \textit{Transformer} and \textit{BERT}. 
All these workloads benefit mainly from kernel fusion optimization of memory intensive ops, 



\subsection{Comparing with Nimble}

We compare with Nimble for \textit{transformer}, 
which is one of the most important structure in modern machine learning models.
Table \ref{tbl:breakdown-nimble} shows the performance breakdown of compute intensive ops, memory intensive ops and CPU time. 
Note that we use library call of cuDNN/cuBLAS for compute intensive ops in Nimble implementation instead of kernel tuning, 
as we do not find open sourced code of the schedules for compute intensive ops described in Nimble report.

It shows that \NAME achieves 2.61$\times$ speedup for memory intensive ops only, 
which is one of the main reason of the overall speedup.
The advantage of \NAME is that, 
it collects shape hints from both shape propagation and shape constraints 
to help with efficient fusions. 
The shape-adaptive fusion configuration strategy further contributes to the overall performance.

Another advantage of \NAME is low-overhead runtime flow. 
Table \ref{tbl:breakdown-nimble} shows that the CPU time with \NAME is only 36.6\% of that with Nimble. 
A small portion of the CPU time reduction comes from reduced kernel launches, 
as \NAME shows a slight reduce of total kernels (table \ref{tbl:kernel-count-nimble}). 
While the main reason comes from that, 
\NAME generated runtime flow works more efficiently with co-optimization of host and device control flow.

\begin{table}
\begingroup
\caption{Performance breakdown for Transformer.}
\label{tbl:breakdown-nimble}
\begin{center}
\begin{small}
\renewcommand{\arraystretch}{0.85}
\begin{tabular}{ccccc}
\hline
Backend    & Comp. bound & Mem. bound & CPU    & E2E    \\ \hline
Nimble     & 66.58       & 56.09      & 65.83  & 188.5  \\
\NAME      & 59.68       & 21.52      & 24.08  & 105.28 \\ \hline
\end{tabular}
\end{small}
\end{center}
\endgroup
\end{table}

\begin{table}
\begingroup
\caption{Kernel number breakdown for Transformer.}
\label{tbl:kernel-count-nimble}
\begin{center}
\begin{small}
\renewcommand{\arraystretch}{0.85}
\begin{tabular}{cccc}
\hline
Backend    & Comp. bound & Mem. bound & Total \\ \hline
Nimble     & 5232        & 8632       & 13924 \\ 
\NAME      & 4476        & 6186       & 10734 \\ \hline
\end{tabular}
\end{small}
\end{center}
\endgroup
\end{table}

\subsection{Gap to Static Optimization}

\NAME can fall back to static compiler automatically for better performance. 
To evaluate the performance of dynamic compiler with static compiler, 
we disable the fall back function and compare the performance between static and dynamic compilers 
with static input for 3 typical workloads.

It shows that \NAME achieves 85\% performance in average 
comparing with static optimization, ranging from 74.5\% to 91.4\%. 
One reason of the gap is that it lacks some fusion optimization opportunity(such as more aggressive graph optimization, fusion decision, and codegen strategy, etc) without shape information, 
even though we have already collected shape hints with \NAME.
\begin{figure}
    \includegraphics[scale=0.45]{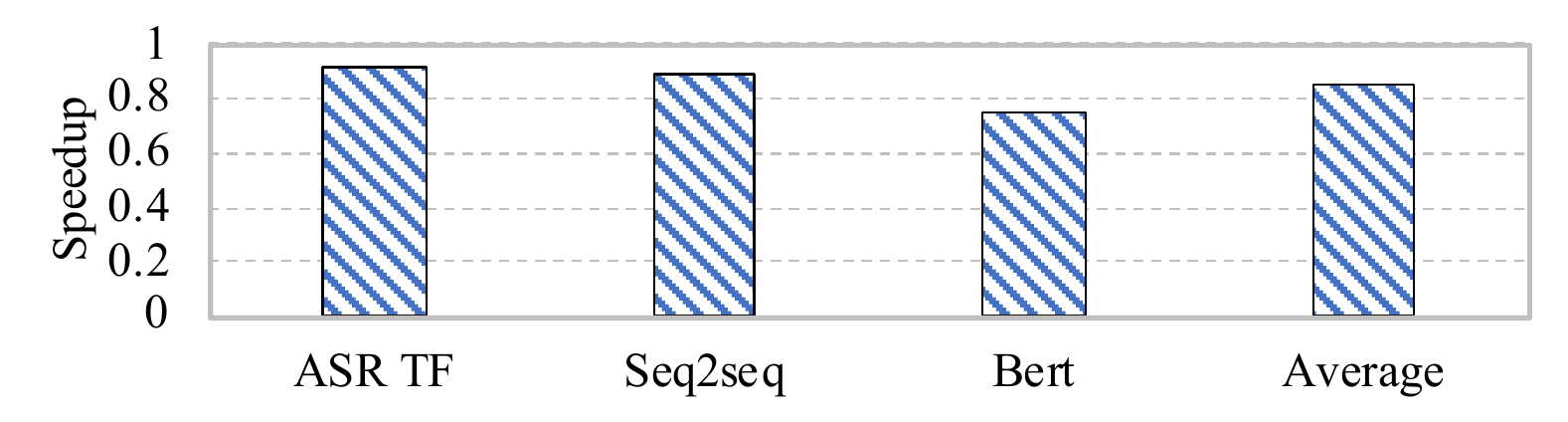}
    \caption{Performance gap to static optimization.}
    \label{fig:performance-gap}
\end{figure}

\section{Related Work}
\label{sec:related}

There are many works that apply kernel fusion optimization for 
small sized kernels in machine learning workloads. 
XLA\cite{xla} fuses kernels just-in-time with a set of rules for ops including element-wise and reductions. 
FusionStitching\cite{zheng2020fusionstitching} expands the scope that 
JIT fusion can target with intermediate value reusing between ops.
Some works\cite{chen2018tvm,zheng2020ansor,vasilache2018tensor,baghdadi2019tiramisu} 
that mainly targets large compute intensive ops 
also have ability of fusion for small kernels.
These techniques are served for static shape scenarios, 
and suffer from severer compilation overhead for dynamic shape workloads.

Lazy compilation\cite{raimandali,nvtfguide} can be applied 
to reduce compilation overhead if unknown shapes are limited. 
However, it loses partial opportunity of kernel fusion optimization, 
and is infeasible to be applied when there are too many unknown shapes.

Nimble\cite{shen2020nimble} addresses the dynamic shape problem by 
building a compiler system based on TVM. 
It proposes a VM approach to interpret dynamic shape processing flow at runtime. 
Instead, \NAME generates runtime flow at compile time to avoid interpretation overhead, 
and exposes more opportunity of host-device co-optimization. 
Meanwhile, \NAME pays more attention to memory intensive fusion comparing with Nimble.

IREE\cite{iree} is an open source MLIR-based end-to-end compiler that lowers ML models to a unified IR optimized for real-time mobile/edge inference against heterogeneous hardware accelerators. IREE provides flexible deployment solutions for the compiled ML models while it is still in its early phase.
\section{Conclusion}

\NAME addresses the dynamic shape optimization problem.
It demonstrates how to build a compiler system based on MLIR infrastructure.
\NAME supplements \textit{HLO} and forms \textit{DHLO}, 
which is a fully dynamic shape representation. 
\textit{DHLO} is served as a hub IR that supports versatile machine learning frameworks. 
The runtime flow, including shape inference, buffer management and host-side control,
is generated by the compiler. 
This is a new attempt that aims to reduce interpretation overhead 
and enrich host-device-joint optimization opportunity.
With shape propagation and shape constraints collecting, 
\NAME applies efficient kernel fusion optimization without full shape information. 
Experiments show that \NAME outperforms state-of-the-art solutions with 1.8$\times$ speedup.

\label{sec:conclusion}

\bibliographystyle{ACM-Reference-Format}
\bibliography{refs}


\end{document}